\documentclass[aps,prl,twocolumn,showpacs,preprintnumbers,amsmath,amssymb]{revtex4}
\usepackage{graphicx}
\usepackage[rightcaption]{sidecap}

\begin{document}

\title{Non-Gaussian fluctuations in stochastic models with absorbing barriers}

\author{Claudia Cianci}
\affiliation{Dipartimento di Sistemi e Informatica, University of Florence, via di Santa Marta 3, 50139 Firenze, Italy, CSDC, INFN}

\author{Francesca Di Patti}
\affiliation{Dipartimento di Fisica, Sapienza Universit\`{a} di Roma, P.le  A. Moro 2, 00185 Roma, Italy}

\author{Duccio Fanelli}
\affiliation{Dipartimento di Energetica, University of Florence, via di Santa Marta 3, 50139 Firenze, Italy, CSDC, INFN}

\begin{abstract}
The dynamics of a one-dimensional stochastic model is studied in presence of an absorbing boundary. The distribution of fluctuations is  
analytically characterized within the generalized van Kampen expansion, accounting for higher order corrections beyond the conventional
Gaussian approximation. The theory is shown to successfully capture the non Gaussian traits of the sought distribution returning an excellent agreement with the simulations, for {\it all times} and arbitrarily {\it close} to the absorbing barrier. At large times, a compact analytical solution for the distribution of fluctuations is also obtained, bridging the gap with previous investigations, within the van Kampen picture and without resorting to alternative strategies, as elsewhere hypothesized. 
\end{abstract}

\pacs{02.50.-r, 05.10.Gg, 05.40.-a, 05.70.Ln}

\maketitle

Studying the dynamics of large ensemble of interacting entities is a fascinating field of investigations of broad applied and fundamental interest. Atom and nuclei result for instance in extended systems of mutually interacting discrete elements. Similarly, proteins can be ideally imagined as a coherent sea of microscopic actors that e.g. densely populate the inside of the cells. Social and human communities are also examples of systems that display rather intricate dynamics, microscopically governed by the very complex network of interlaced connections among individuals.
Surprisingly, and as follows a widespread observation of cross-disciplinary breath, regular collective modes can spontaneously emerge at the macroscopic level, as resulting from the erratic dynamics of the microscopic discrete constituents \cite{alan,dau1,dau2,bia}.

The time evolution of such inherent stochastic system is generally described in terms of a master equation \cite{gardiner}, a differential equation for the probability of observing the scrutinized system in a certain configuration at given time. In practically all cases of interest, solving the master equation proves a task of formidable complexity, and approximated strategies need to be implemented to gain analytical insight into the system being examined. The celebrated van Kampen expansion \cite{vanKampen} represents a viable technique to enable one for analytical progress. It is customarily believed that such a method works efficiently well provided the system is defined in a open domain or, conversely, if it evolves sufficiently far from any existing boundaries \cite{alan}. As an emblematic example, when the system has to face an asymptotic extinction, thus evolving towards an attractive absorbing state, the van Kampen perturbative scheme is only assumed appropriate for short times. Recently, and to eventually bypass these supposedly stringent limitations, a different perturbative approach has been pioneered in \cite{dipatti} that suites for large times, when the system is feeling the absorbing barrier. Working within this generalized setting, and operating with reference to a paradigmatic model of systems with an absorbing state, the voter model, a closed expression for the distribution of fluctuations was obtained which agrees with direct simulations. The derivation rests however on speculative grounds, which, despite the a posteriori validation, seem to lack of a solid physical interpretation,  fully justified from first principles.

The purposes of this Letter are twofold. On the one side, and with reference to the same version of the voter model
as considered in \cite{dipatti}, we will analytically demonstrate that by extending the van Kampen expansion to include
higher orders corrections \cite{grima}, beyond the classical approximation, allows us to accurately reproduce the observed
distribution of  fluctuations {\it at any time}. Non Gaussian traits reflecting the presence of the absorbing
barrier are nicely captured by the method, which proves therefore accurate {\it also close to the boundary}. Even more
interesting, the van Kampen solution is shown to converge at late times to the distribution calculated in \cite{dipatti},
this latter being hence explained within a sound and universal descriptive picture.

Let us start by introducing the stochastic discrete voter model. As in the spirit of \cite{dipatti}, we consider a system
made of $N$ elements in mutual interactions, possibly organized in different species. Label with $X_1$ the elements of a
specific species and with $X_0$ all the other entities. The following chemical equations are proposed to rule the microscopic
dynamics:
\begin{equation*}
X_{1} + X_{0}  \stackrel{1}{\longrightarrow}  2X_{0},
\end{equation*}
\begin{eqnarray*}
X_{0} + X_{1} & \stackrel{1-\nu}{\longrightarrow} & 2X_{1},\\
X_{1} + X_{1} & \stackrel{\nu}{\longrightarrow} & X_{1}+X_{0}
\end{eqnarray*}

The master equation which stems from the above system reads:
\begin{eqnarray}
\frac{d}{dt}P_{n}(t)&=&(\epsilon_{n}^{-}-1)[T(n+1|n)P_n(t)]\nonumber \\
&\phantom{=}&+(\epsilon_{n}^{+}-1)[T(n-1|n)P_n(t)]\label{master}
\end{eqnarray}
where $P_n(t)$ is the probability of photographing the system at time $t$ in a configuration with
$n$ individuals belonging to the population of $X_1$ and $\epsilon_{n}^{\pm}$ are  the step operators \cite{vanKampen}.
The transition rates are given by:
\begin{eqnarray*}
T(n+1|n) & = &(1-\nu)\frac{(N-n)}{N}\frac{n}{N}\\
T(n-1|n) & =&\frac{n}{N}\frac{N-n}{N}+\nu\frac{n}{N}\frac{n}{N}
\end{eqnarray*}
where the initial states are the right entries and the final states the left ones. As follows the above,
$n=0$ is an absorbing state while $n=N$ corresponds to a reflecting barrier. The van Kampen approach requires
imposing:
\begin{equation}\label{ip}
\frac{n}{N}=\phi(t)+\frac{\xi}{\sqrt{N}}
\end{equation}
where $1/\sqrt{N}$ plays the role of a small parameter and paves the way to the
perturbative expansion hereafter discussed. By inserting the working ansatz
(\ref{ip}) into the master equation (\ref{master}),
and hierarchically organizing the resulting terms with respect to
their $N$-dependence, one obtains at the first order the mean-field deterministic
equation for the continuum concentration $\phi(\tau)$ ($\tau$ being the rescaled time $t/N$), namely $
d \phi/d\tau = -\nu\phi $, whose solution reads $\phi(\tau) =\phi_0\exp(-\nu\tau)$. Higher order
contributions results in a generalized Fokker-Planck equation for
the new probability $\Pi(\xi,\tau)=P\left(\phi(\tau)+\xi/\sqrt{N},\tau \right)$. By truncating the expansion at the second order yields the standard
Fokker-Planck equation, which predicts Gaussian fluctuations.
Allowing instead for higher order corrections, generates a cascade of
terms whose relative weights are controlled by the finite size $N$.
After a lengthy algebraic derivation one ends up with:
\begin{eqnarray}
\label{eq:FP}
\frac{\partial\Pi}{\partial\tau}&=&\sum_{k=1}^{\infty}\frac{1}{(k+1)!}\frac{1}{N^{(k-1)/2}}\frac{\partial^{k+1}}{\partial\xi^{k+1}}\Big[f(\phi,k+1)\Pi\Big]
\nonumber\\
&+&\sum_{k=1}^{\infty}\frac{1}{k!}\frac{1}{N^{(k-1)/2}}\frac{\partial^k}{\partial\xi^k}\Big[g(\phi,\xi,k)\Pi\Big] \label{fpgen}\\
&+&\sum_{k=3}^{\infty}\frac{1}{(k-1)!}\frac{1}{N^{(k-1)/2}}\frac{\partial^{k-1}}{\partial\xi^{k-1}}\Big[q(\xi^2,k-1)\Pi\Big]\nonumber
\end{eqnarray}
where:
\begin{displaymath}
f(\phi,k)=
\left\{
\begin{array}{ll}
2\phi-2\phi^2-\nu\phi+2\nu\phi^2 & \quad \text{for } k  \text{ even}\\
\nu\phi & \quad \text{for } k  \text{ odd}
\end{array} \right .
\end{displaymath}
\begin{displaymath}
g(\xi,\phi,k)=
\left\{
\begin{array}{ll}
2\xi-4\phi\xi+4\nu\phi\xi-\nu\xi & \quad \text{for } k  \text{ even}\\
\nu\xi & \quad \text{for } k  \text{ odd}
\end{array} \right .
\end{displaymath}
\begin{displaymath}
q(\xi,k^2)=
\left\{
\begin{array}{ll}
2\xi^2(\nu-1) & \quad \text{for } k  \text{ even}\\ 
0 & \quad \text{for } k  \text{ odd}
\end{array} \right .
\end{displaymath}

Formally, the positiveness of the probability $\Pi(\cdot)$ is not guaranteed a priori under the
generalized Fokker-Planck evolution, an observation that was made rigorous in \cite{pawula1,pawula2}. However,
with reference to specific case studies \cite{riskenVollmer}, it  was shown that
unphysical negative values are just occasionally attained by $\Pi(\xi,\tau)$,
and punctually localized in the tails of the distribution. The phenomenon fades off when including a sufficiently
large number of terms in the development. The adequacy of the prediction can be a posteriori evaluated via a
direct comparison with the numerical experiments.

\begin{SCfigure*}[][tb]
\begin{tabular}{cc}
\includegraphics[scale=0.3]{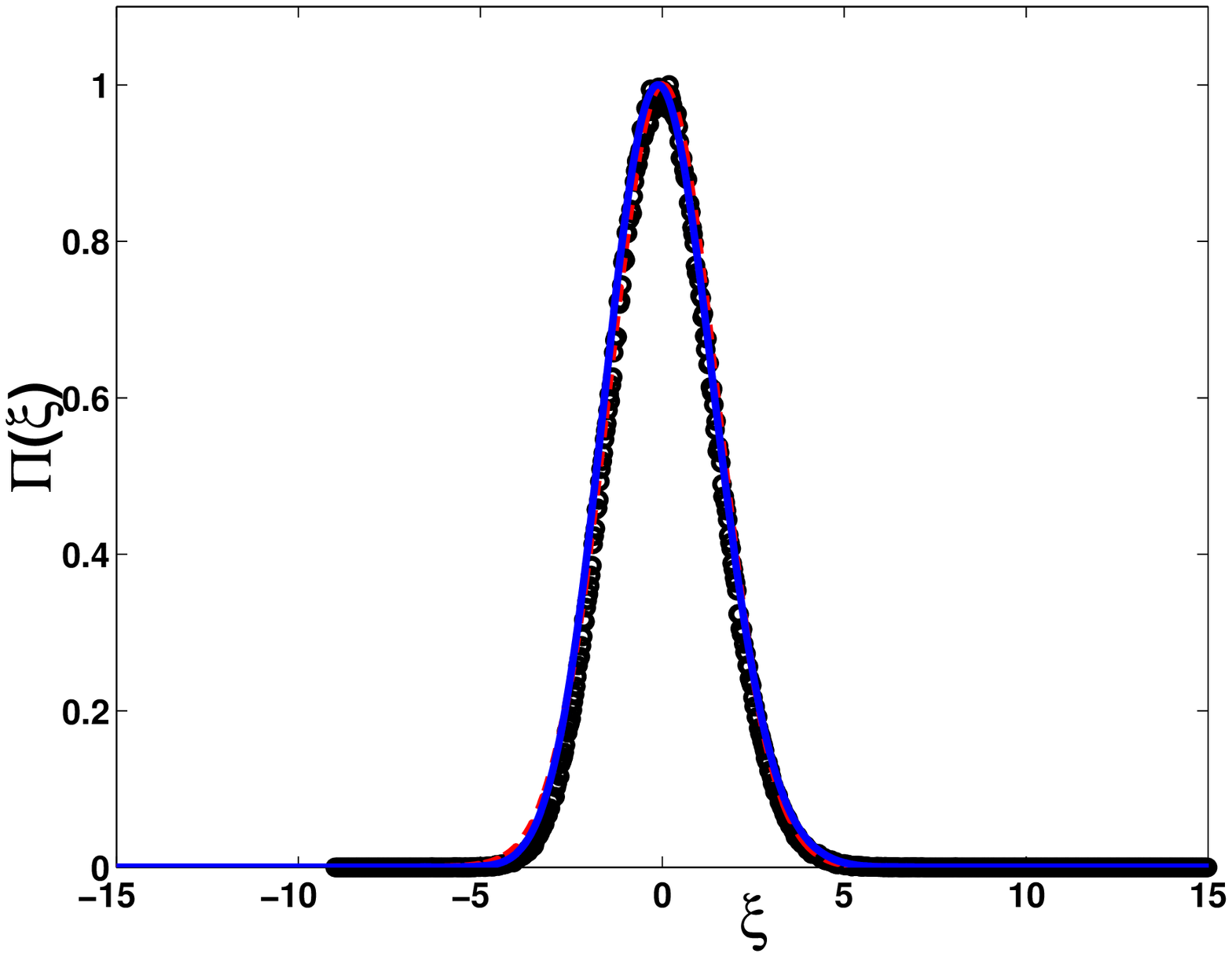}&
\includegraphics[scale=0.3]{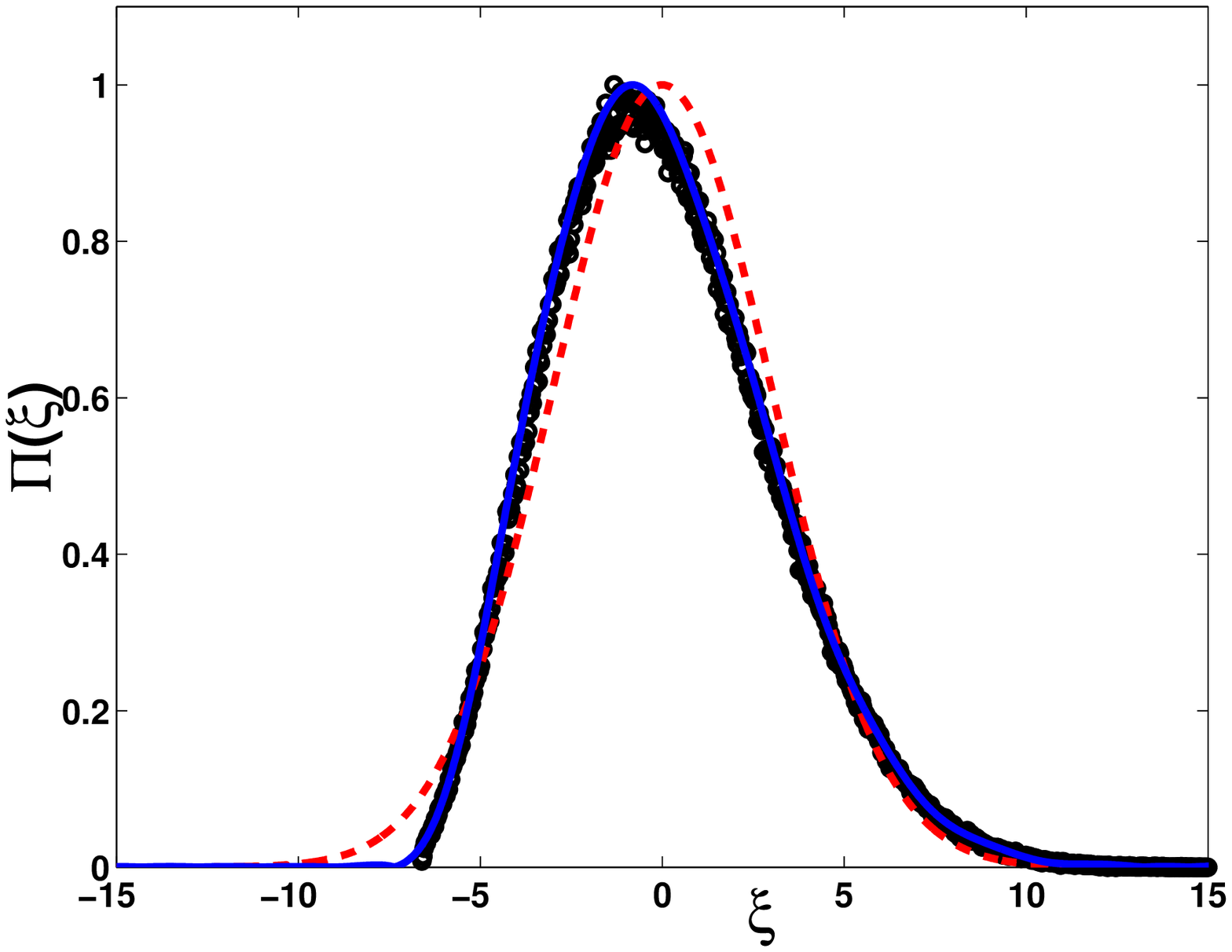}\\
(a) & (b)\\
\includegraphics[scale=0.3]{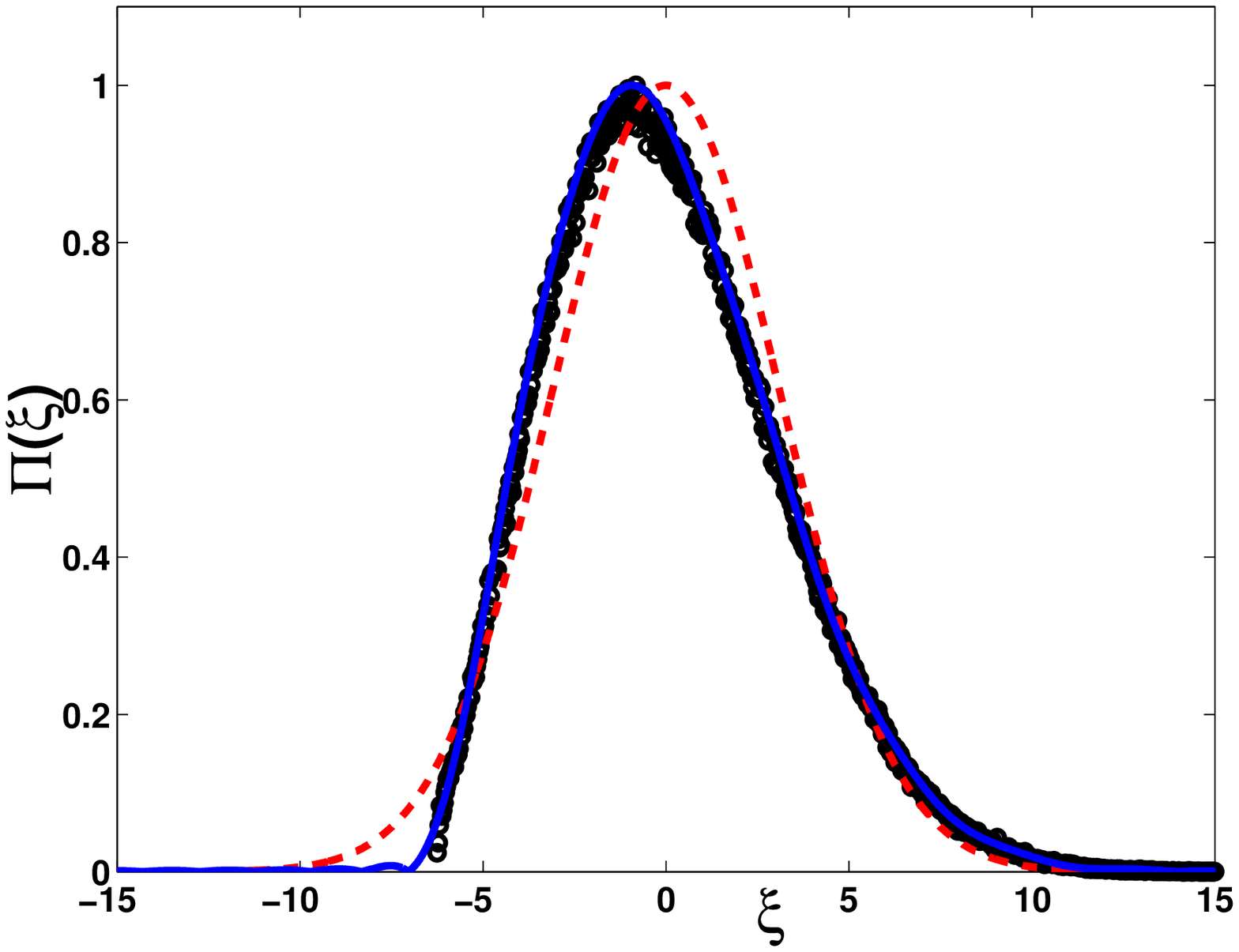}&
\includegraphics[scale=0.3]{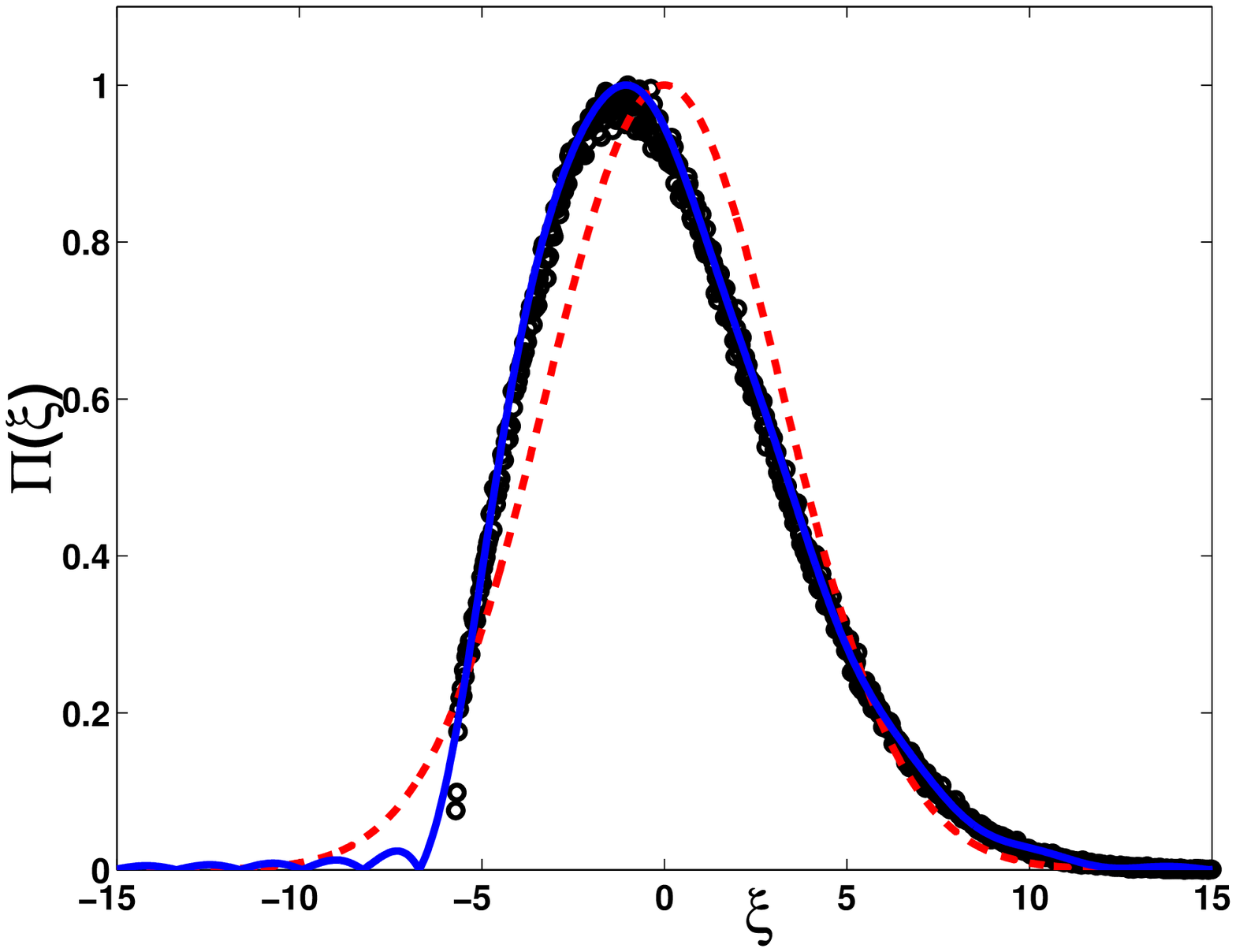}\\
(c) & (d)
\end{tabular}
\caption{The distribution of fluctuations at distinct rescaled times $\tau$. The snapshots refer to: (a)
$\tau=5$, (b)  $\tau=36$, (c)  $\tau=41$ and (d)  $\tau=50$. The symbols stand for direct stochastic simulations. The solid lines represent the theoretical predictions as obtained within the generalized Fokker Planck scenario. We have in particular truncated the sums in the Fokker-Planck (\ref{eq:FP}) to $k=3$ ($1/N^{3/2}$ corrections) and included $200$ moments in the final estimates of the generating function. The dashed lines refer to the Gaussian solutions obtained working within the van Kampen expansion at the, conventional, next to leading approximation ($1/N^{1/2}$ terms). Here $\nu=0.01$ and the distributions are normalized so to have the maximum equal to one.\label{fig:simulations}}
\end{SCfigure*}

To progress with the calculation, we set off to estimate the moments of the sought distribution  $\Pi(\xi,\tau)$. Let
us recall that the moment of order $h$ is defined as
\begin{equation*}
\langle\xi^{h}\rangle=\int_{-\infty}^{\infty}\Pi(\xi)\xi^{h}d\xi.
\end{equation*}
Multiply both sides of the generalized Fokker-Planck
equation  by the factor $\xi^h$ and integrate over $\mathbb{R}$ in
$d\xi$. A straightforward manipulation yields to:
\begin{eqnarray}
\label{momenti}
&&\frac{d}{d\tau} \langle\xi^h  \rangle= \nonumber\\
&&+\sum_{k=1}^{h-2}\frac{1}{(k+1)!}\frac{f(\phi,k+1)}{N^{(k-1)/2}}\frac{h!(-1)^{k+1}}{(h-(k+1))!} \langle\xi^{h-(k+1)}\rangle \nonumber \\
&&+\sum_{k=1}^{h-1}\frac{1}{k!}\frac{g(\phi,\xi,k)}{N^{(k-1)/2}}\frac{h!(-1)^{k}}{(h-k)!} \langle\xi^{h-k+1} \rangle  \label{eq:momentsDE} \\
&&+\sum_{k=3}^{h+2}\frac{1}{(k-1)!}\frac{q(\xi^2,k-1)}{N^{(k-1)/2}}\frac{h!(-1)^{k-1}}{(h-(k-1))!} \langle\xi^{h-(k-1)+2} \rangle \nonumber
\end{eqnarray}
where use has been made of the supposed regularity of the distribution $\Pi(\xi,\tau)$ to drop out the boundary terms
resulting from integrating by parts \footnote{In the definition of the moments we  integrate from $-\infty$ to $\infty$. In principle, the maximum extent of the allowed (negative) fluctuations is limited by the presence of the absorbing boundary. Strictly,
the lower extreme of the above integral should read $-\sqrt(N) \phi(\tau)$. Assuming however that the distribution of
fluctuations $\Pi(\xi,\tau)$ is exactly zero at $\xi=-\sqrt(N) \phi(\tau)$, one recovers the same equations for
the moments as those obtained by formally extending the domain of integration to $-\infty$.}.

We therefore dispose of a closed system of first oder differential equations for the moments of the distribution $\Pi(\xi,\tau)$. We can
integrate it numerically and so estimate the quantities $\langle\xi^h\rangle$, for all $h$, at any time $\tau$. The knowledge of the moments enables us to immediately reconstruct the characteristic function, and so recover, upon Fourier transform inversion, the distribution $\Pi(\xi,\tau)$. The predicted profiles are displayed in Fig. \ref{fig:simulations}  (solid line) for different times. A comparison is drawn with the outcome of direct stochastic simulations based on the exact Gillespie algorithm \cite{gillespie} (symbols), returning excellent agreement. The distribution of fluctuations displays clear non-Gaussian traits. It gets in fact more and more skewed as time progresses, reflecting the non trivial interplay with the absorbing boundary. Surprisingly, and at odds with what customarily believed, the van Kampen ansatz proves accurate well beyond the Gaussian 
approximation that is often invoked to justify its intrinsic validity. As a corollary, it seems tempting to argue that the transformation (\ref{ip}) from discrete to continuum variables is an {\it exact one}, and not just an approximation that presumably descends from the central limit theorem, as occasionally speculated.

It is also very instructive to analyze the asymptotic fate of the distribution of fluctuations, as predicted within the realm of the van Kampen theory. Based on intuition, we expect that when time goes to infinity, the distribution $\Pi(\xi, \tau)$ converges to a Dirac delta centered in zero. Indeed, plugging into the moments' equations (\ref{fpgen}), the asymptotic mean-field solution $\phi =0$, and looking for stationary solutions of the obtained system (i.e. setting the derivatives to zero), one readily gets $\langle\xi^h\rangle=0$ $\forall h$, the moments of a delta function. However, for times large enough that $\phi \simeq 0$, but before the system has relaxed to its stationary state, the generalized Fokker-Planck equation (\ref{eq:FP}) reads:
\begin{equation}
\label{FPdipattiN}
\frac{\partial\Pi}{\partial \tau}  = \nu \frac{\partial}{\partial\xi}\left(\xi \Pi \right) + 
\frac{2-\nu}{2 \sqrt{N}} \frac{\partial^2}{\partial^2 \xi}\left(\xi \Pi \right)
\end{equation}
where we have only retained the term in $1/\sqrt{N}$ dropping higher orders corrections. Perform now the scaling
$\xi \rightarrow \xi' / \sqrt(N)$. The equation (\ref{FPdipattiN}) can be cast in the form:  
\begin{equation}
\label{FPdipatti}
\frac{\partial \Pi}{\partial \tau}  = \nu \frac{\partial}{\partial\xi'}\left(\xi' \Pi \right) + 
\frac{2-\nu}{2} \frac{\partial^2}{\partial^2 \xi'}\left(\xi' \Pi \right) \quad .
\end{equation}

The large time distribution $\Pi(\xi', \tau)$ is therefore insensitive to the system size $N$ and bears consequently universal
traits. Equation (\ref{FPdipatti}) can be solved analytically (see also \cite{dipatti}) to give:
\begin{eqnarray}
\Pi(\xi', \tau)  &=& \frac{2 \nu}{2-\nu} \frac{1}{1-e^{-\nu \tau}} \exp\left[  
\frac{2\nu (\xi'+ \xi'_0 e^{-\nu \tau})}{(2-\nu)(1-e^{-\nu \tau})}\right] \nonumber \\
&\times& \left( \frac{\xi'}{\xi'_0} e^{\nu \tau} \right)^{-\frac{1}{2}} I_1 \left(  
\frac{4 \nu \sqrt{\xi'_0 \xi' e^{\nu \tau}}}{(2-\nu)(e^{\nu \tau}-1)} \right) \label{FPsolution}
\end{eqnarray}
where $I_1(\cdot)$ is the modified Bessel function of the first kind. For large $\tau$, recalling that $I_1(x) \simeq x/2$ when $x$ is small, one can approximate Eq. (\ref{FPsolution}) as:
\begin{equation}
\label{FPsolution1}
\Pi(\xi', \tau) \propto \xi_0' \left( \frac{2 \nu}{2-\nu } \right)^2 \exp(-\frac{2 \nu}{2-\nu} \xi') \exp(- \nu \tau) 
\end{equation} 

\begin{figure*}[tb]
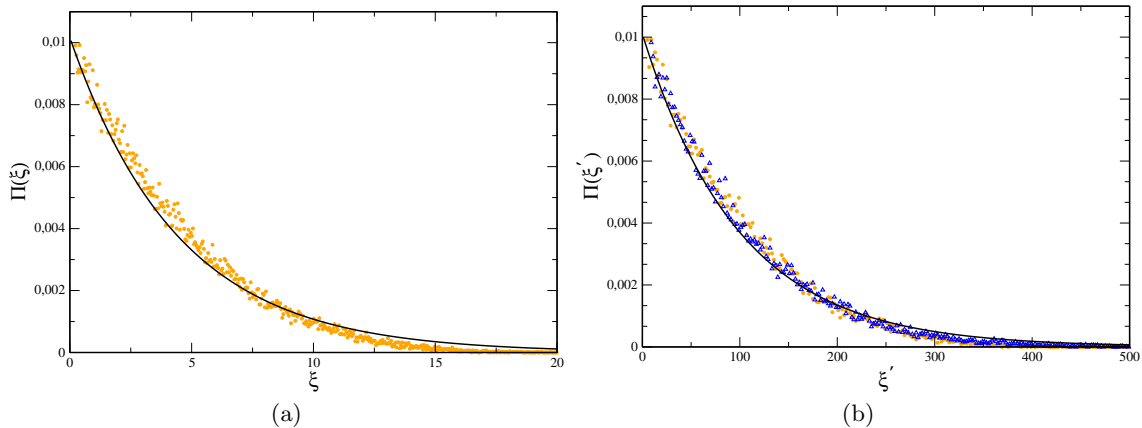

\begin{tabular}{cc}
\includegraphics[scale=0.3]{PI1.eps}&
\includegraphics[scale=0.3]{PI2.eps}\\
(a) & (b)
\end{tabular}
\caption{The distribution of fluctuations at large  times. Left panel: the distribution $\Pi(\xi, \tau)$ is plotted versus $\xi$ at $\tau=388$. Symbols refer to the simulations ($N=1000$), while the solid line stands for the (normalized) solution (\ref{FPsolution1}) after the change of variable $\xi' \rightarrow \xi \sqrt(N)$ is performed. Right panel: the distribution $\Pi(\xi', \tau)$ is plotted as function of the rescaled $\xi'$. Symbols refer to numerical simulations relative to distinct $N$. In particular, $N=500$ (circles) and $N=1000$ (triangles). The solid line stands for the (normalized) solution (\ref{FPsolution1}). Here $\nu=0.01$ and the distributions are normalized to unit.\label{fig2}}
\end{figure*}

Operating with the rescaled variable $\xi'$, which, it is worth emphasizing, emerges naturally within the van Kampem expansion, when the large time limit is being considered, it is equivalent to inserting into the governing master equation the modified ansatz $n = N \phi + \xi'$. This latter corresponds to the strategy adopted in \cite{dipatti} for the specific choice $\alpha=0$. In other words, and interestingly enough, the expected fluctuations  $\xi'$ are comparable to the discrete population size $n$, when the absorbing boundary is being approached. We have therefore recovered exactly the same solution as obtained in \cite{dipatti}, while working within the generalized, but conventional, van Kampen approach. The adequacy of (\ref{FPsolution1}) is challenged in Fig. \ref{fig2} versus numerical simulations returning a perfect quantitative agreement. Notice that different distribution profiles recorded at distinct values $N$, nicely superpose when the rescaled fluctuations $\xi'$ is employed. 

As a side remark, we stress that the same conclusion can be drawn working in the Fourier space and operating under analogous approximations.
Retaining only  $1/\sqrt{N}$ corrections in (\ref{momenti}), assuming $\phi \rightarrow 0$ and performing the scaling $\langle\eta^h\rangle=\langle\xi^h\rangle N^{\frac{h}{2}}$\footnote{This is equivalent, in Fourier space, to the transformation $\xi \rightarrow \xi' / \sqrt(N)$ of the original fluctuation $\xi$.}, one immediately obtains the following $N$ independent differential equations for the moments evolution: 
\begin{equation}
\label{Momentseta}
\frac{\partial}{\partial\tau}\langle\eta^h\rangle=-\nu h \langle\eta^h\rangle+\frac{h(h-1)}{2}(2-\nu)\langle\eta^{h-1}\rangle.
\end{equation}

Equations (\ref{Momentseta}), here obtained within the extended van Kampen scenario, could be also derived  via the alternative, supposedly distinct, approach discussed in \cite{dipatti}. Equations (\ref{Momentseta}) can be in fact straightforwardly deduced from Eq. (\ref{FPdipatti}) (see also Eq. (8) in \cite{dipatti}) following the same strategy for the evaluation of the moments as outlined above.

In conclusion, the van Kampen approximation works effectively for {\it all times}, well beyond the Gaussian approximation and in a regime where the presence of the absorbing boundary is definitely important. The method returns in fact the correct asymptotic solution (a delta function centered in the origin), but also converges to the large time solution calculated in \cite{dipatti}, which is therefore contextualized within a general 
descriptive picture. At moderate times, after the Gaussian approximation has broken down, direct comparisons with the numerical experiments, as reported in this Letter, testify on the excellent predictive ability of the van Kampen theory. This is an important observation, that will certainly motivate using the van Kampen machinery beyond the limited domains of applications for which it was originally conceived, and/or later referred to \cite{dipatti}. Future investigations will be targeted to extending the current analysis to cases where the absorbing boundary competes with a non trivial stable fixed point, as well as models for which the notion of space  is explicitly accounted for \cite{dau2,lugo}. 
\bibliographystyle{apsrev4-1}
\bibliography{bibliography.bib}
\end{document}